\documentclass[twocolumn,showpacs,preprintnumbers,amsmath,amssymb]{revtex4}
\usepackage{graphicx}
\usepackage{dcolumn}
\usepackage{bm}
\usepackage[english]{babel}
\begin{document}

\title{Analysis of patterns formed by two-component diffusion limited aggregation}

\author{E.\;B.\;Postnikov}%
 \email{postnicov@gmail.com}
\affiliation{Kursk State University, Department of Theoretical
Physics, Radishcheva st., 33, 305000 Kursk, Russia}%
\author{A.\;B.\;Ryabov}%
 \email{a.ryabov@icbm.de}
\affiliation{University of Oldenburg, ICBM, 26111 Oldenburg, Germany}%
\author{A.\;Loskutov}%
 \email{loskutov@chaos.phys.msu.ru}
\affiliation{Moscow State University, Physics Faculty,
119992 Moscow, Russia}%

\begin{abstract}

We consider diffusion limited aggregation of particles of two different kinds.
It is assumed that a particle of one kind may adhere only to another particle
of the same kind. The particles aggregate on a linear substrate which consists
of periodically or randomly placed particles of different kinds. We analyze the
influence of initial patterns on the structure of growing clusters. It is shown
that at small distances from the substrate, the cluster structures repeat
initial patterns. However, starting from a critical distance the initial
periodicity is abruptly lost, and the particle distribution tends to a random
one. An approach describing the evolution of the number of branches is
proposed. Our calculations show that the initial patter can be detected only at
the distance which is not larger than approximately one and a half of the
characteristic pattern size.

\end{abstract}

\pacs{61.43.Hv}

\keywords{Diffusion-limited aggregation, diffusion-annihilation, wavelets}

\maketitle

\section{Introduction}

It is well known that diffusion limited aggregation (DLA), firstly proposed by
Witten and Sander \cite{Witten81}, is a quite suitable model of a number of
physical phenomena. For example, it describes kinetic processes of the growth
of electro-deposited dendrites, colonies of bacteria, viscous fingers in fluid
mixtures, etc. Such a variety of applications induced theoretical and numerical
analysis of DLA, and at present it is a well-studied model which has been
verified on various theoretical and experimental levels (from the mean-field
approach to the detailed microscopic representation). A comprehensive review
can be found, for instance, in the book \cite{Meakin}.

Generalization of DLA can provide a proper description of more complex pattern
formation processes (see, e.g., \cite{Roeder93, Jensen94}). This interest was
inspired by the emergence of micro- and nanoscale structures, which are built
by deposition of diffusing particles on a surface. In particular, the authors
of the paper \cite{Meyer01} experimentally studied layer-by-layer deposition
and investigated the cluster structure in each layer. Moreover, there are important
biophysical and biotechnological examples: the formation of self-assembled
patterns such as biomimetics and biominerals \cite{Murr05, Imai07};
technologies of fabrication and usage of templates for the cell/protein
aggregate formation \cite{Yousaf01, Turner05, Khademhosseini06}, etc. Finally,
the formation of clusters on a given pattern can serve as a tool for the
amplification of scattered or emitted signals \cite{Suci08}.

A natural generalization of the classical DLA is a multicomponent cluster
formation. In this case the growing structure is created from distinct kinds of
particles which interact with each other in different ways. For the
multicomponent DLA, random deposition of particles on the growing cluster
changes the particle distribution in layers so that the initial pattern becomes
more and more fuzzy with the distance from the initial germ. This induces
questions about the influence of the deposition on the structure of layers and
about the competing roles of the growth and diffusion processes in the
formation of multilayer clusters.

The DLA-model of two non-interacting kinds of particles was advanced in the
paper \cite{Nagatani91}. The authors showed that the branches of two kinds can
coexist and unrestrictedly grow if and only if the particles of different kinds
appear in the system with the equal probability. A more detailed description of
the aggregation of various kinds of particles was proposed in \cite{Tchijov96,
Rodriguez05}, where it was argued that if the branches of various kinds are
impenetrable then the structure of aggregates is completely determined by
initial fluctuations. Recently it has been shown \cite{Postnikov07} that the
fractal dimension of a multicomponent DLA-cluster can be found by means of the
coarse-grained mean-field approach. This method allows  correct estimations
of the fractal dimension even if we neglect the fluctuations in the
distribution of particles.

In the present paper we consider two-component DLA on a substrate (germ) which
consists of periodically or randomly distributed particles. We assume that new
particles of both kinds arise with the same probability and can aggregate only
on the particles of the same kind. The number of initial single-kind patches on
the substrate determines the maximal number of branches which can appear in the
formed cluster. However, is the substrate contains short patches then in a few
neighbor layers initial periodicity is lost, and we observe a random particle
distribution. Otherwise, if the initial characteristic size of patches is large
enough then the pattern periodicity persists until a certain layer. In the
further layers the periodic structure becomes more and more fuzzy and the
particle distribution tends to a random one.

To analyze the scaling properties of this process we used two characteristics:
the number of branches at a certain distance from the substrate and the
spectrum of the wavelet transform with Haar basis. Both approaches show
self-similarity of the clusters. The wavelet analysis shows that at the
distance larger than approximately 3/2 of the initial patch size the branch
distribution becomes indistinguishable from a random one.

\section{The Model}

\begin{figure}
\centering
\includegraphics[width=\columnwidth]{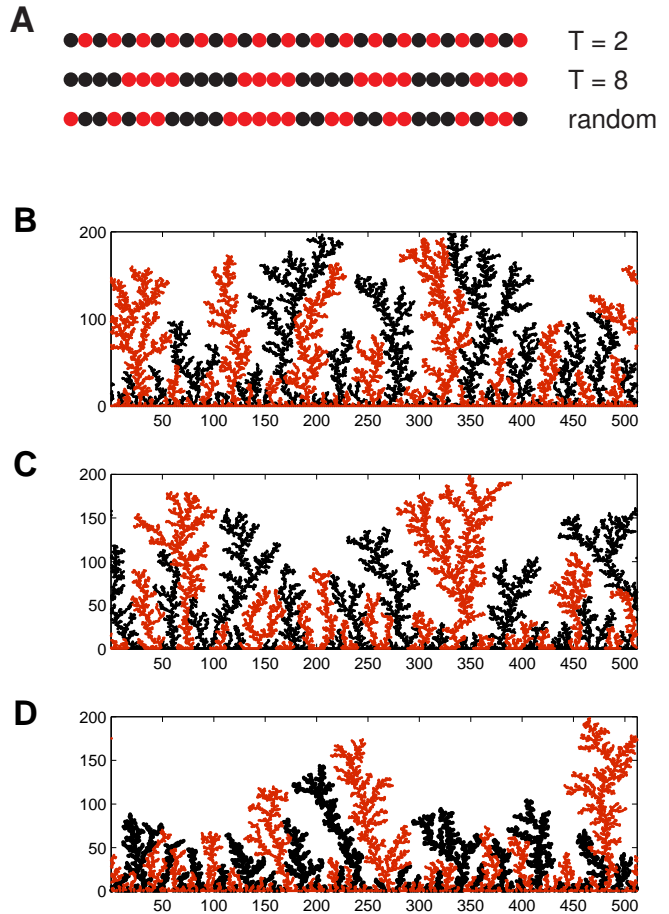}
\caption{(Color online) (A) --- Periodically and randomly assembled substrates
of particles of two non-interacting kinds (grey and black/red and black -- in color online version). (B), (C), and (D)
--- Typical DLA-clusters on different substrates: $T=2$, $T=32$, and random
initial distribution, respectively. To be more visible, width $W$ of the
initial substrate consists of $512$ particles (i.e. 1/8 used for numerical
simulations.} \label{fig_clusters}
\end{figure}

As a model of layer-by-layer random deposition on a given initial matrix, we
consider a two-component off-lattice DLA process in cylindrical geometry.
Particles of two non-interacting kinds (marked as grey and black (red and black in color online version), see
Fig.~\ref{fig_clusters}) randomly walk and aggregate on initial substrate in a
rectangular area of height $H$ and width $W$ with lateral periodic boundary
conditions.

An initial pattern (Fig.~\ref{fig_clusters}A) is formed by particles of two
kinds, which are periodically or randomly distributed along the bottom side of
the rectangle. A new particle of a randomly chosen kind appears at the upper
side of the rectangular area and diffuses within it. If this particle
leaves the area across the upper side, then we generate a new one. Colliding with a
seed of another kind, the particle is reflected and proceeds its motion. Upon a
collision with a seed of the same kind, it sticks and becomes a part of the
cluster. After aggregation a new particle appears at the 
upper side and so on. As a result, branches of different kinds grow on the initial
substrate (Figs.~\ref{fig_clusters}B-\ref{fig_clusters}D). To avoid the
domination of particles of one kind \cite{Nagatani91}, we assume that the
appearance of particles of every kind is equiprobable.

In our simulation, the height $H$ of the rectangle is equal to the fourfold
maximal cluster height (which is $500$ particle diameters), and the width $W$
is equal to $4096$ particle diameters. Also, to enhance our simulations we used
a few improvements of the algorithm. Firstly, the particle wandering within a
``fjord'' composed of particles of another kind may dramatically increase the
simulation time. To avoid this we limit the particle lifetime by $50000$ steps.
%
%
 Secondly, we dynamically change the step size of the random walk, increasing it
far from clusters and reducing it up to $0.1$ of the particle diameter in the
vicinity of clusters \cite{about_algorithm}.
%
%
To control the finite
size effects we repeated the same simulations on two times shorter substrates
($W=2048$). These experiments show the same results until the number of cluster
brunches is larger than 10. Thus before this threshold the influence of finite
size effects is negligible.

To analyze the evolution of the branch distributions we simulate the growth of
clusters on random and periodical substrates (Fig.~\ref{fig_clusters}A). In the
former case particles of both kinds are initially randomly distributed on the
line. In the latter case, they are placed periodically so that $N$ particles of
one kind alternate with $N$ particles of another kind. Thus, the period of such
patterns is $T = 2N$. We perform simulations for the periods $T = 2$, $4$, $8$,
$16$, $32$, and $T=64$ particles.

The limitation in the period length ($T\leq 64$) is determined by the finite
size of substrates and periodic boundary conditions. This periodicity implies
that the actual number of branches will always be  integer.

Thus our results are valid until the number of branches is essentially larger
than one. In fact, see below, it is possible to neglect the boundary effects
until the cluster includes at least 10 branches.

For numerical analysis, we discretized the distance from the substrate using the
particle diameter as a  minimal unit of length. Thus, we define a set of
horizontal layers with the height of one unit, which are numbered in accordance
with their distance from the substrate. The substrate is located in
the first layer.

\section{Cluster morphology}

Apparently, with an increase of the distance from the substrate, the influence
of initial patterns vanishes, and the initially periodic branch distribution
becomes more and more random. Figs.~\ref{fig_clusters}B--\ref{fig_clusters}D
show that the initial size of mono-color patches on the substrate plays a
crucial role for the cluster morphology. If a substrate has a short period ($T
\le 4$) then the initial periodic structure is rapidly lost. In fact, for $T=2$
the distribution of colors appears to be random already in the second layer.
However, if $T \ge 8$ then a DLA-cluster can develop inside of each mono-color
region, preserving its initial periodicity until a certain layer.

The average mass of cluster layers gives a rough characteristic of clusters on
different initial substrates. On random and short-periodic ($T \le 4$)
substrates the mass of layers decays as a power law $M(n)\propto n^{\alpha_m}$,
where $\alpha_m = -0.39$, thereby resulting in the fractal dimension $D\simeq
1.61$ of these clusters.

\begin{figure}[h!]
\includegraphics[width=\columnwidth]{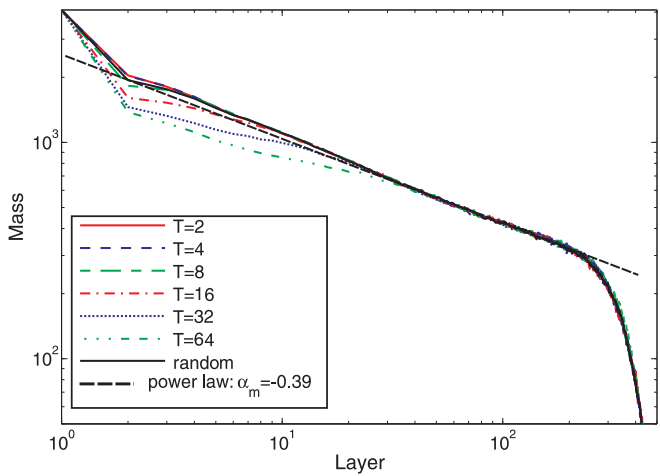}
\caption{(Color online) The layer mass as a function of the layer number for
clusters growing on periodically and randomly assembled substrates.}
\label{fig_mass}
\end{figure}

This value is close to the fractal dimension of multicomponent DLA clusters in
radial geometry  \cite{Postnikov07}.  As shown in this article, an additional
screening by branches of different kinds decreases the fractal dimension in
comparison with the standard value  $1.78$.

By contrast, the long-periodic $(T \ge 8)$ substrates lead to sparser
aggregates. The main reason is that every mono-color patch on the substrate
gives rise to a single DLA branch which, attaching almost all particles of the
same kind, shades its base. Therefore, the wider the initial mono-color
patches, the smaller the density of the cluster arising on this substrate. Note
that from a certain layer the layer mass approaches that obtained for randomly
assembled substrates (black line in Fig.~\ref{fig_mass}).

More information about the cluster morphology we can obtain evaluating the
number of different branches. For instance, in radial geometry this number
increases with the distance from the cluster center \cite{Mandelbrot02}. By
contrast in cylinder geometry, the number of branches decreases with the layer
number, because branches of the same kind can merge, stopping growth of other
branches squeezed between them. To evaluate the number of branches, we trace
the colors of particles along each layer and register changes of the colors.
Thus, we combine two branches of the same kind into one if there is no branch
of another kind between them in a certain layer. In other words, we 
assume that
these branches merge. This procedure is correct in the layers where the
aggregation has already stopped, however it underestimates the number of
branches for those regions of cluster which actively adsorb new particles.

\begin{figure}[h!]
\includegraphics[width=\columnwidth]{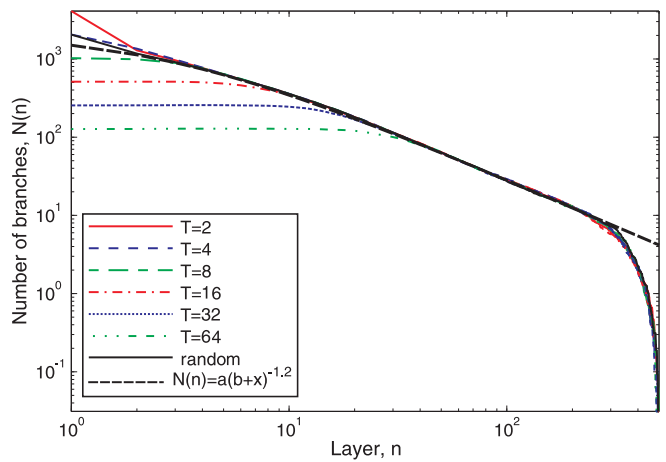}
\caption{(Color online) The number of branches as a function of the layer
number in clusters growing on periodically and randomly assembled substrates.
The black dashed line shows the power low $N(n) = a (b + x)^{-1.2}$, where
$a=7288$, $b=2.74$. } \label{fig_crossections}
\end{figure}

Fig.~\ref{fig_crossections} shows the average number of branches as a function
of the layer number. As expected, for all initial distributions this value
decreases. However, we want to highlight several aspects. Firstly, all curves
finally approach the curve obtained on the random substrate. The number of
branches, $N$ decreases with the layer number, following the power low
$N(n)\propto (n + b)^{-1.2}$. The initial shift $b$ probably occurs because
initially the particles are distributed without gaps between them and the
transition to a sparse structure takes a few layers.

Secondly, the numbers of branches on a random and on a periodic with $T=4$
substrate almost coincide. This coincidence apparently arises from the fact
that initially these substrates have the same average number ($1024$) of
different mono-color patches (see Appendix \ref{ap:rand}). Thirdly, if a
substrate has more patches than a random one, then the number of branches
abruptly decreases up to the characteristic value obtained on the random
substrate.

For instance, if $T=2$ then each patch consists of a single particle and the
number of branches decreases threefold already in the second layer. Aggregates
on substrates with a large initial period ($T \ge 8$) have less branches and
preserve their periodicity approximately until they have as many branches as on
a random substrate. After this layer, the branches begin to extinct and their
total number decreases and coincides with the number of branches on randomly
assembled substrates. Therefore, the number of branches growing on a random
substrate defines a natural limit, and aggregates preserve their initial
periodicity as long as they have less branches than clusters growing on a
random substrate at the same layer.

\section{Wavelet spectral analysis}

The number of branches provides just a rough insight into the cluster
structure. More detailed information can be gained from the spectral analysis.

Divide every layer into $W$ square cells of a unit size. In each cell of the
layer $n$ we define the piecewise-constant function $f_n(x)$ such that $f(x)$
is equal to $+1$ or $-1$ if the center of a grey (red in color online version) or  black particle,
respectively, is located within the cell, and $f_n(x) = 0$ otherwise. Note that
sometimes in our simulations a cell can confine the centers of two particles.
To resolve this uncertainty we used the color of the left particle. However,
these events rarely occurred, and thus they can not influence the final
result \cite{rare_events}.

Since the piecewise-constant function $f_n(x)$ takes discrete values
$\{-1, 0, 1\}$ on a uniform grid of cells, for the spectral analysis it is
natural to use the wavelet transform with Haar orthonormal basis functions
$$
\psi_{jk}(x)=\sqrt{2^j}\psi(2^j x - k) \ ,
$$
where $j$ and $k$ are the integer numbers, which characterize the scale
and the shift, respectively, of Haar wavelet basis function
$$
 \psi  (x) = \left\{ {\begin{array}{*{20}r}
 {1,}  \\
 { - 1,}  \\
 {0,}  \\
 \end{array}} \right.\begin{array}{*{20}l}
 {x \in [0,1/2)}  \\
 {x \in [1/2,1]}  \\
 {x \notin [0,1] \ .}  \\
 \end{array}
$$
Every layer includes $W$ cells. Thus, to represent the function $f_n(x)$, the
largest wavelet pattern should have the scaling factor $J = \log_2 W$, and the
wavelet expansion of the function $f_n(x)$ reads
\begin{equation}
  f_n(x) = c_0+
   \sum_{j = 0}^J  \sum_{k = 0}^{W-1}
    d_{jk}^{\ (n)}\psi _{jk} (x) \ ,
    \label{Haar_expansion}
\end{equation}
where the index $n$ refers to the layer number, $c_0$ is the mean value of the
function $f_n(x)$ and the coefficients
$$
d_{jk}^{\ (n)}= \int_0^W f_n(x) \psi_{jk}(x) dx
$$
show the partial impact of the wavelet with period $T=2^j$ in the sites from
$k$ to $k + T$. Thus, the sum of the wavelet components with the same scaling
factor,
\begin{equation}
F^{\ (n)}_j(x)=\sum_{k = 0}^{W-1} d^{\ (n)}_{jk}\psi_{jk}(x) \ ,
\label{wvl_harm}
\end{equation}
gives the total impact of wavelets with period $T=2^j$ and represents a
counterpart of an individual harmonic in the Fourier expansion. Exploiting
this analogy, we can refer to the sum
\begin{equation}
E^{\ (n)}_j =  \sum_{k = 0}^{W-1} \left(d^{\ (n)}_{jk}\right)^2
\label{Haar_spectrum}
\end{equation}
as a power of the spectral component including all wavelets with period $T =
2^j$. Thus, the set of values $\{ E^{\ (n)}_j \}$ where $j=0,1, \dots, J$
represents the global power spectrum of the wavelet transform of the layer $n$.

The power spectrum of the initial substrates with period $T$ has only a single
non-zero component $E^{\ (1)}_{j^*}$, where $j^* =\log_2 T$. This component
represents the main harmonic of the initial wavelet spectrum. Other components
appear in the second layer, and their relative impact increases with the layer
number, reflecting the transition from the periodic to the random structure of
clusters.

However, during this transition the value $E^{\ (1)}_{j^*}$ dominates, thereby
reflecting the fact that initial periodicity is preserved in the cluster
structure, i.e., the distribution of branches ``remembers'' the initial
distribution.

Consider now the normalized average power of the main harmonics:
\begin{equation}
\overline{E}_{j^*}(n)=\frac{1}{\sqrt{2^{j^*-1}} N_T} \left\langle
E^{\ (n)}_{j^*} \right \rangle \ ,
\label{main_harmonic}
\end{equation}
where $\langle \ldots \rangle$ denotes the average over all realizations. To
compare the power of different harmonics on the same scale, we introduced the
specific renormalization factor, $1\Big/\left(\sqrt{2^{j^*-1}} N_T\right)$.
This factor compensates for the differences in the number of initial periods,
$N_T=W/T$, and in the scales of the basis wavelet functions with various $j^*$.

\begin{figure}
\includegraphics[width=\columnwidth]{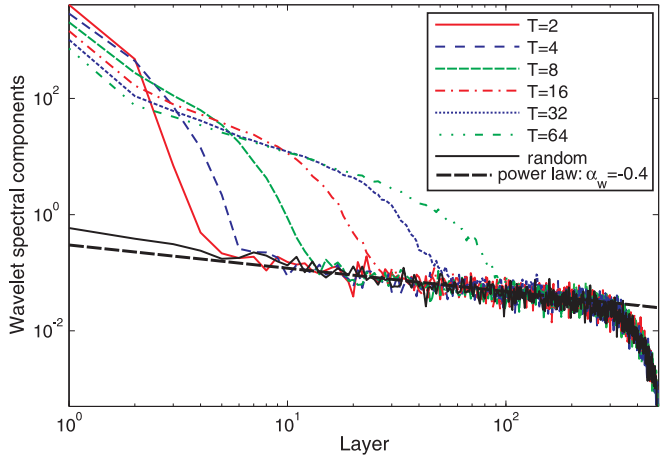}
\caption{(Color online) The distribution of normalized global power spectrum of
wavelet components corresponding to the various initial periods.}
\label{fig_wavelets}
\end{figure}

Fig.~\ref{fig_wavelets} shows the normalized power of the main harmonics for periodic
substrates (color solid lines) in comparison with one harmonic
($T=2$) on randomly assembled substrates (dash-dotted line).

On randomly assembled substrates the power spectrum equally includes all
possible periodic components, which from the outset follow the power law with
the slope $\alpha_w = 0.4$ (see 
dash-dotted black line in Fig.~\ref{fig_wavelets}).

Consider how the power of the main harmonics evolves with layer number for
substrates which have large initial periodicity: $T\ge 8$. Until the number of
branches is less than the number of branches growing on a random substrate
(compare with Fig.~\ref{fig_crossections}), the power of the main harmonics
decreases relatively slow. However, at a critical layer, when this number is
exceeded, this decrease becomes much steeper and the power of the main
harmonics abruptly declines to the level of other harmonics. Finally, the power
of all harmonics goes over the same power law with the exponent $\alpha_w =
-0.4$.

Note that our approach has a few common traits to the method based on averaged
wavelet coefficients, which was suggested in \cite{Simonsen98} for the analysis
of self-affine time series. However, in contrast to the this work, we analyze
not the inner structure of a unique function but the set of functions
corresponding to a certain layer of a cluster.

Thus, we can select three stages. During the first stage the clusters preserve
their initial periodicity and the initial harmonics prevails. During the second
stage the main harmonics abruptly decreases and some branches rapidly extinct.
In the last stage the distribution of branches becomes indistinguishable from a
random distribution. If the initial period is small ($T \le 4$) then we observe
only the last two stages.

Note that because of the finite cluster height we observe this behavior only
approximately within the first 300 layers. Afterwards all harmonics abruptly
decay. Furthermore, the substrate width seemingly 
influences the development of clusters with initial period $T=64$.

\section{Mean-field annihilation \label{MFA}}

Initially the substrate has an invariance with respect to a shift by one period
length. A particle, which randomly attaches to a branch,  limits access of
other particles to the adjacent branch and breaks the local symmetry. This
process is stochastic and local changes accumulate and finally lead to
extinction of some branches. A a result, the influence of initial patterns
reduces with distance from substrates.

A rough insight into the dynamics of this process can be gained by a mean-field
approximation arising from the simple diffusion-annihilation model suggested in
\cite{Redner97}. Namely, if two branches of kind $A$ shade a branch of kind $B$
between them, then we can write this as the annihilation process
$$
A_n + B_n + A_n = A_{n+1} \ ,
$$
where $n$ is the layer number.

Let $N(n)$ be the average number of branches of the same kind, and let $k$ be
the annihilation reaction rate of the branches. The inverse concentration $1/c$
determines the mean distance between two branches of the same kind. Therefore,
the characteristic time after which two branches can merge and shade another
branch is equal to $\Delta n \propto 1/kc$. Furthermore, the decrease of the
concentration should be proportional to the density of branches. Therefore, the
differential equation of this annihilation process has the form:
$$
\dot{N}\cong \frac{\Delta N}{\Delta n}= -kN^2.
$$
Solving this equation we obtain
\begin{equation}
N(n)=\frac{1}{k}\left(\frac{1}{N(0)k}+n\right)^{-1}.
 \label{PDE}
\end{equation}

This solution qualitatively describes the dependence of the number of brunches
in Fig.~\ref{fig_crossections}, it also predicts the initial shift $b=1/N(0)k$.
Note that the exponent of this dependence is integer, which is common for
simple mean-field approximations. However, as the two exponents ($-1$ and
$-1.2$) are sufficiently close, this formula gives a good approximation in a
certain range.

\section{Discussions and Conclusion}

In this paper we analyze the structure of DLA-clusters, growing in a
rectangular area with lateral boundary conditions. The clusters are built up by
aggregation of particles of two different kinds on random or periodic
substrates. Growing DLA branches can merge with each other, stopping the growth
of other branches between them. This decreases the number of distinguishable
branches and changes the structure of the cluster as a whole. Thus, since this
process is stochastic, the influence of initial patterns reduces with distance
from substrates.

In more details, we investigated how the rate with which the system loses its initial
periodicity depends on the distance from the substrate and showed that there
are three characteristics areas. In the first area, which includes first few
layers, the average number of clusters preserves and the wavelet harmonics
corresponding to the initial periodicity decay slowly. This area exists only if
the number of different patches on the substrate is initially smaller than that
on a randomly assembled one. In the following layers (where the number of
different branches becomes close to that on a random substrate, or in the
second layer if this number was exceeded initially) the initial periodicity
drastically decreases and the distribution of branches approaches to a random
one. Finally, the average number of branches in a layer coincides with the
number of clusters which appear on a randomly assembled substrate, and their
distribution becomes random. In this area the evolution of the number of
branches can be estimated within the mean-field approach.

\begin{figure}
\includegraphics[width=\columnwidth]{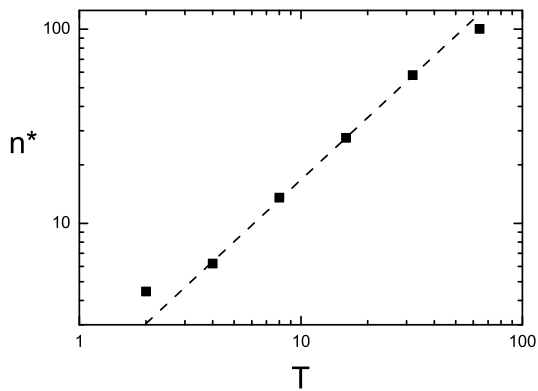}
\caption{The maximal distance $n^*$ from  the initial substrate after which the
initial periodicity cannot be detected in dependence of the initial period. The
black line is fitting $n^*=1.46 T^{1.06}$. } \label{fig_crossover_points}
\end{figure}

From the practical point of view, our results allow to elaborate a criterion
for the validation of experimental methods for the selection and amplification
of micropatterns via layer-by-layer deposition of specific reagents (see, e.g.,
\cite{Suci08}). Fig.~\ref{fig_wavelets} shows that for every initial period of
substrates there is a maximal layer number, $n^*$, after which the distribution
of branches loses initial periodicity and the magnitude of all harmonics
becomes equal. Fig.~\ref{fig_crossover_points} shows that this value scales
with the size of initial patterns. For linear fitting we exclude $T=2$, because
this distribution loses its regularity almost in the fist layer, and $T=64$,
because in this case $n^*$ is close to the layers where we observe effects of
finite cluster height. All other points follow a power low dependence with the
exponent $\gamma=1.06$. Accuracy of our simulations do not allow to confirm
that $n^*$ grows faster than $T$ ($\gamma>1$). Thus as an approximate
estimation we suggest $n^*<3T/2$. This dependence shows the principal limit for
detection of the substrate structure at a certain distance from the substrate.

\appendix

\section{Number of branches in random distribution
\label{ap:rand}}

Consider a randomly assembled substrate. The sort of a particle in each
location is randomly assigned with the probability $1/2$. It is evident that
the same kind for $n$ successive particles will be assigned with the
probability
$$
P_n = \frac{1}{2^n} \ .
$$
Thus, the average size of initial clusters of this kind is
$$
S = \sum_{n=1}^{\infty} n P_n =  \sum_{n=1}^{\infty}  \frac{n}{2^n} = 2 \ .
$$
Therefore, the average number of clusters on the random substrate should be the
same as we have in the periodic pattern, when $2$ particles of one kind succeed
$2$ particles of another kind, i.e. for $T=4$.

\bibliography{mDLA}

\begin{thebibliography}{21}
\bibitem[{abo()}]{about_algorithm}
\bibinfo{note}{The step size of the random walk can be chosen dynamically
  \cite{Menshutin06}. According to this approach, if there are no particles
  within a circular of radius $R$, then the step size equals $R - d$, where $d$
  is the particle diameter. Otherwise (to avoid missing the cluster particles)
  we use the step size of 0.1 of the particle diameter. In both cases the step
  direction is given by uniformly distributed random numbers $\phi \in [0, \ 2
  \pi]$. We also cheked that two- and fourfold change of the minimal step size
  and the maximal life time of a particle leads us to the same scaling laws.}
\bibitem[{rar()}]{rare_events}
\bibinfo{note}{This effect is important only when two particles of different
  colors are located within one cell. However, this event is extremely rare,
  because usually there is an essential gap between branches of different
  color. We have detected only two such events in our simulations, therefore
  their impact is negligible and cannot affect our results.}
\end{thebibliography}

\end{document}